\documentclass[twocolumn]{aastex631}
\usepackage{multirow}
\usepackage{subfigure}
\usepackage{amsmath}

\begin{document}

\title{Joint observations of late universe probes: cosmological parameter constraints from gravitational wave and Type Ia supernova data}

\author{Jie Zheng}
\affiliation{Gravitational Wave Observatory, Henan Academy of Sciences, Zhengzhou, 450046, China}

\author{Xiao-Hui Liu}
\affiliation{National Astronomical Observatories, Chinese Academy of Sciences, Beijing, 100101, China}
\affiliation{School of Astronomy and Space Science, University of Chinese Academy of Sciences, Beijing, 100049, China}

\author{Jing-Zhao Qi}
\affiliation{Department of Physics, College of Sciences, MOE Key Laboratory of Data Analytics and Optimization for Smart Industry, \\
Northeastern University, Shenyang, 110819, China}

\correspondingauthor{Jing-Zhao Qi}
\email{qijingzhao@mail.neu.edu.cn}

\begin{abstract}
The growing tensions between the early Universe and the late Universe increasingly highlight the importance of developing precise probes for late cosmology. As significant late-Universe probes, Type Ia supernovae (SNe Ia) and gravitational waves (GWs) can provide measurements of relative and absolute distances, respectively. Their complementary nature is likely to break the degeneracies among cosmological parameters, thereby yielding more precise constraints. In this study, we use 43 gravitational-wave sources from the Third LIGO–Virgo–KAGRA Gravitational-Wave Transient Catalog (GWTC–3) and 1590 SNe Ia from Pantheon+ compilation to constrain the dark energy models, as an attempt to achieve precise late-Universe cosmological constraints. For the dark siren GW event, we estimate the corresponding redshift using the binary black hole redshift distribution model. The combination of GW and SNe Ia data could provide the precision on the Hubble constant $H_0$ and the present matter density $\Omega_{m}$ of approximately 20\% and 8\% for the $\Lambda$CDM model. If we consider the equation of state of dark energy $w$, the combination sample constrain the precision of $w$ to approximately 30\%. Although the combination of GW and SNe Ia observations effectively breaks degeneracies among various cosmological parameters, yielding more stringent constraints, the precision of these constraints still does not meet the stringent standards required by precision cosmology. However, it is reasonable to anticipate that, in the near future, the joint observations of GWs and SNe Ia will become a powerful tool, particularly in the late Universe, for the precise measurement of cosmological parameters.

\end{abstract}


\section{Introduction}
\label{sec:intro}
The Hubble constant $H_0$ characterizes the Universe's current expansion rate and provides significant insights into our understanding of the Universe's behavior.
However, the discrepancies in the measurements of the Hubble constant obtained from different observations have led to the well-known Hubble tension.
Many analyses have pointed out that the Hubble tension is not simply attributed to systematic errors in the observations \citep{2016PhLB..761..242D, pantheon,2020arXiv200710716E,riess}.
There are also significant disagreements arising in other cosmological parameters as inferred from early- and late-Universe probes.
For instance, the curvature parameters $\Omega_{K}$ inferred from CMB and baryon acoustic oscillation (BAO) measurements are in disagreement with each other \citep{2020NatAs...4..196D,2021PhRvD.103d1301H}.
Additionally, within $\Lambda$CDM model, the parameters $S_{8}\equiv \sigma_{8}\sqrt{\Omega_{m}/0.3}$, where $\sigma_{8}$ is the root-mean-squared of matter fluctuations on a $8h^{-1}$Mpc scale and $\Omega_{m}$ is the current matter density, inferred from CMB indicates a 2-3$\sigma$ disagreement with that deduced from weak lensing survey \citep{2012MNRAS.427..146H,2020NatAs...4..196D,2021A&A...646A.140H,2021ApJ...908L...9D,2022PhRvD.105b3520A}.

The tensions between early and late Universe observations are interesting because, if not caused by systematic errors, they suggest the presence of new physics beyond the current $\Lambda$CDM model framework \citep{2019NatAs...3..891V}. To confirm our previous understanding of the Universe's evolution, we need to use new and independent late-Universe probes for cosmological analysis, as early-Universe probes, while highly precise, are model-dependent. Furthermore, a single late-Universe probe is not enough for achieving precision cosmology, as each probe has its own strengths and weaknesses, with different sensitivities to various combinations of cosmological parameters and physical processes. Therefore, the most effective way to address the crisis arising from the inconsistencies between early and late Universe observations is to combine different late-Universe probes, taking advantage of their complementary nature.

The successful detection of gravitational waves (GWs) has opened a new window to explore the accelerated Universe \citep{BBH,151226,170104,170814,170608,170817}, showing the potential for understanding the inconsistency between the measurements of the early and late Universe \citep{Qi:2020rmm,Cao:2021zpf,Hou:2022rvk}. As first pointed out by \citet{Schutz1986}, GWs generated from compact binary coalescence can be standard sirens, since the strain amplitude encodes the absolute distance from the source. However, the challenge for standard siren cosmology is to identify the redshifts of GW sources.

There are usually three primary mechanisms to standard siren cosmology: "bright siren," "dark siren," and "spectral siren." The GW signal, which provides the distance to the host galaxy, while its electromagnetic (EM) counterpart provides the redshift information, can be referred to as a bright siren \citep{2005ApJ...629...15H,PhysRevD.74.063006,PhysRevD.77.043512,2010ApJ...725..496N,2018Natur.562..545C}, like GW170817 event \citep{170817}. If the EM counterpart is unavailable, but a complete galaxy catalog is available, one can consider all galaxies within the GW localization volume as potential host galaxies, which are often referred to as dark sirens \citep{Schutz1986,PhysRevD.86.043011,2018Natur.562..545C,Gray2020PRD,ligod}. Another method is to use known features in the source population to obtain the redshift and distance from the GW signal alone, which can be called spectral sirens \citep{BBH,2021PhRvD.104f2009M,ligod,PhysRevLett.129.061102}.

These approaches are useful for exploring the role GWs play in measuring the Hubble constant \citep{Qi:2021iic,Zhang:2019ylr,Wang:2019tto,Zhao:2019gyk,Jin:2020hmc,Jin:2022qnj,Zhang:2019loq}. However, it is known that GW observations only weakly constrain other cosmological parameters \citep{20171005}. In this scheme, determining $\Omega_{m}$ and the equation of state (EoS) of dark energy $w$ requires other reliable cosmological probes.

SNe Ia have long been the ideal standard candle, since its peak luminosity is strictly correlated with its absolute luminosity, and the absolute luminosity is obtained in terms of the absolute magnitude $M_{B}$ of SNe Ia \citep{1993ApJ...413L.105P}. Since $M_{B}$ is degenerate with $H_0$, using SNe Ia to measure $H_0$ is theoretically debatable, but it can constrain $\Omega_{m}$ and $w$ precisely.
Thus, using the combination of GWs and SNe Ia is exciting for cosmology, since they are completely independent and complementary, providing more robust constraints on the cosmological parameters that describe the expansion history of the Universe.

In this paper, we aim to investigate the potential of two emerging cosmological probes, GWs and SNe Ia, in constraining cosmological parameters, particularly in light of the inconsistencies between measurements of the early and late Universe. Our study focuses on assessing the effectiveness of cosmological inference using these probes, their capacity to break parameter degeneracies, and their potential contributions to a more comprehensive understanding of the Universe's evolution. To achieve this, we employ 43 compact binary coalescence candidates, including 42 binary black hole (BBH) merger events and a binary neutron star (BNS) merger event, GW170817 from the third LIGO-Virgo-KAGRA GW transient catalog (GWTC-3) \footnote{https://zenodo.org/records/8177023}, and the "Pantheon+" sample \citep{pantheonp}, removing the data points in redshift range $z < 0.01$. We use these datasets to constrain the cosmological parameters of two well-established cosmological models: the $\Lambda$CDM model and the $w$CDM model. In Sec.~\ref{sec:2}, we present the methodology of measuring the cosmological parameters, and the data used in this work. We show our results and discussion in Sec.~\ref{sec:3}. Finally, the main conclusions are summarized in Sec.\ref{sec:4}.

\section{Data and Methodology}
\label{sec:2}

\subsection{Gravitational Wave Data}


GWTC-3 contains 90 compact binary coalescence candidate events with at least a 50\% probability of being astrophysical in origin. Notably, 47 of these GW events have a network matched filter signal-to-noise ratio (SNR) greater than 11 and an Inverse False Alarm Rate (IFAR) higher than 4 years.
Following the work of \citet{LIGOScientific:2021aug}, we select 42 BBH events and a BNS event, GW170817 \citep{170817}. We exclude one BNS event GW190425, two neutron star-black hole (NS-BH) merger events GW200105 and GW200115, and one asymmetric mass binary event GW190814.
The reason for excluding these events is that their EM counterparts have not been detected and the equation of state of neutron star is undetermined, making it impossible to obtain their redshift information.
For the GW170817 event, which has confirmed electromagnetic (EM) counterparts, the host galaxy and its redshift can be determined directly. However, for the BBH events lacking confirmed EM counterparts, alternative techniques must be employed to infer the source redshift, such as the population method \citep{mastro37,2019ApJ...883L..42F,2021PhRvD.104f2009M,PhysRevLett.129.061102} and the galaxy catalog method \citep{2012PhRvD..86d3011D,2018ApJ...863L..41F,Gray2020PRD,2021JCAP...08..026F}.


In this work, we employ the population method to infer the cosmological parameters for several reasons. Firstly, the population method avoids introducing extra observational errors and the incompleteness of galaxy catalogs, as we focus on investigating the potential of GW and SNe Ia data in constraining cosmological parameters in this work. Secondly, the galaxy catalog method generally requires fixing the BBH population mass models, which introduces significant biases if the population assumptions are incorrect \citep{BBH,LIGOScientific:2021aug}. Although \citet{2023JCAP...12..023G} has improved this issue, it is crucial to recognize that the likelihood of the out-of-catalog term remains influenced by the population assumptions. In this work, we employ the population method to estimate both cosmological and population parameters.


The GW signal directly provides the redshifted detector-frame masses $m_{1,d}$, $m_{2,d}$, which are related to the source-frame masses $m_{1}$, $m_{2}$ by
\begin{equation}
\label{eq:mi}
m_{i}=\frac{m_{i,d}}{1+z}.
\end{equation}
The key point of the dark siren analysis is to break the degeneracy between the source-frame mass $m_{i}$ and the source redshift $z$. 
This requires the knowledges of the mass distribution of compact objects and their merger rate model. 

With the observations of the luminosity distance of the source $D_{L}$, the detector-frame masses $m_{1,d}$, $m_{2,d}$, the sky position $\Omega$, and the spin parameters $\chi$, the rate of compact binary coalesces (CBC) mergers can be modeled as a function \citep{2023arXiv230517973M}:
\begin{equation}
    \frac{\mathrm{d} N}{\mathrm{~d} D_L \mathrm{~d} \Omega \mathrm{~d} m_{1,d} \mathrm{~d} m_{2,d} \mathrm{~d} \chi \mathrm{d} t_d},
\end{equation}
where $t_{d}$ is the detector unit time. In practice, we marginalize over the sky position $\Omega$ and the spin parameters $\chi$, as they are not used in breaking the mass-redshift degeneracy. 
The detector events parameters and the source events parameters can be denoted as $\theta_{\mathrm{D}}=\{ D_{L}, m_{1,d}, m_{2,d} \}$ and $\theta_{\mathrm{S}}=\{z, m_{1}, m_{2}\}$, respectively. 
As a reminder, we need to work with a detector rate written in terms of detector-frame variables $\theta_{\mathrm{D}}$, with the aim of inferring properties in source-frame variables $\theta_{\mathrm{D}}$. 
This coordinate transformation can be expressed as \citep{2023arXiv230517973M}
\begin{equation}
\begin{split}
\frac{\mathrm{d} N}{\mathrm{d} \theta_{\mathrm{D}} \mathrm{d} t_{d}}&=\frac{\mathrm{d} N}{\mathrm{~d} \theta_{\mathrm{S}} \mathrm{d} t_{s}} \frac{\mathrm{d} t_{s}}{\mathrm{d} t_{d}} \frac{1}{\mathrm{det} J_{\mathrm{D} \rightarrow \mathrm{S}}} \\
&=\frac{\mathrm{d} N}{\mathrm{d} \theta_{\mathrm{S}} \mathrm{d} t_{s}} \frac{1}{1+z} \frac{1}{\operatorname{det} J_{\mathrm{D} \rightarrow \mathrm{S}}},
\end{split}
\end{equation}
where the term $\frac{1}{1+z}$ encodes the clock difference between the source-frame and detector-frame, $\operatorname{det} J_{\mathrm{D} \rightarrow \mathrm{S}}$ is the determinant of the Jacobian from the change of variables $\theta_\mathrm{D} \rightarrow \theta_\mathrm{S}$, written as 
\begin{equation}
\label{eq:17}
\operatorname{det} J_{\mathrm{D} \rightarrow \mathrm{S}}=\frac{\partial d_L}{\partial z}(1+z)^2,
\end{equation}
where the factor $(1+z)^2$ comes from the transformation from source-frame to detector-frame of both $m_{1}$ and $m_{2}$, and 
\begin{equation}
    \frac{\partial d_L}{\partial z}=\frac{D_{L}(z)}{1+z}+\frac{c(1+z)}{H_{0}}\frac{1}{E(z)},
\end{equation}
where $E(z)=H(z)/H_{0}$. 
In the end, the detector rate model is parameterized as \citep{2023arXiv230517973M}:
\begin{equation}
\begin{split}
\frac{\mathrm{d} N}{\mathrm{~d} \theta_{\mathrm{D}} \mathrm{d} t_d}&= R_0 \psi(z| \Phi) p\left(m_{1}, m_{2} | \Phi\right) \times \\
&=\frac{\mathrm{d} V_c}{\mathrm{~d} z} \frac{1}{1+z} \frac{1}{\operatorname{det} J_{\mathrm{D} \rightarrow \mathrm{S}}},
\end{split}
\label{eq:18}
\end{equation}  
where $\Phi$ represents a set of \textit{population hyper-parameters} that are common to the entire population of GW sources, which contains the distribution of BH sources in the population $\Phi_{m}$, and the cosmological parameters $\Phi_{c}$, here, $R_{0}$ is the CBC merger rate per comoving volume per year, the term $\psi(z | \Phi)$ is a function describing the redshift evolution of BBH merger rate, $p\left(m_{1}, m_{2} | \Phi\right)$ describes the mass distribution of compact objects, and the $dV_{c}/dz$ is the differential comoving volume.

For the redshift evolution of BBH merger rate $\psi(z | \Phi)$, it is common to apply a cosmic star formation rate (SFR) model from \citet{2014ARA&A..52..415M,2020ApJ...896L..32C} to describe it, as the black hole formation rate may track the SFR. This relationship can be characterized by a low-redshift power-law slope $\gamma$, a peak at redshift $z_{\mathrm{p}}$, and a high-redshift power-law slope $k$ after the peak, and 
\begin{small}
\begin{equation}
\psi\left(z | \Phi \right)=\left[1+\left(1+z_{\mathrm{p}}\right)^{-\gamma-k}\right] \frac{(1+z)^\gamma}{1+\left[(1+z) /\left(1+z_{\mathrm{p}}\right)\right]^{\gamma+k}}.
\end{equation}    
\end{small}

The source mass models $p\left(m_{1}, m_{2} | \Phi\right)$ can be factorized as
\begin{equation}
    p(m_{1},m_{2}| \Phi ) = p(m_{1} | \Phi ) p(m_{2} | m_{1},\Phi), 
\end{equation}
where $m_1$ and $m_2$ represents the primary mass and the secondary mass, respectively. $p(m_{1}| \Phi)$ represents the distribution of the primary mass component, while $p(m_{2}| m_{1},\Phi)$ represents the distribution of the secondary mass component given the primary mass.
For the primary mass distribution, we consider three popular phenomenological mass models used in Ref.~\citet{2019ApJ...882L..24A,2021ApJ...913L...7A,LIGOScientific:2021aug}: the \textsc{Truncated} model, the \textsc{Broken power law} model, and the \textsc{Power law + peak} model.
The secondary mass distribution is modeled by a truncated power-law distribution with slope $\beta$ between a minimum mass $m_{\mathrm{min}}$ and maximum mass $m_{1}$.
The expression of the population mass models can be found in appendix. In Tab.~\ref{tab:prior}, we summarize the mass model parameters and their prior distributions used in the Bayesian parameter estimations.

\begin{table*}
    \centering
    \caption{The priors for the population hyper-parameters for the three phenomenological mass models. We adopt the same values as \citet{LIGOScientific:2021aug}}
    \label{tab:prior}
    \begin{tabular}{lll}
        \hline
        \hline
        \multicolumn{3}{c}{\textsc{Truncated}} \\
        \hline
        \textbf{Parameter} & \textbf{Description} & \textbf{Prior} \\
        \hline
        $\alpha$ & Spectral index for the PL of the primary mass distribution. & $\mathcal{U}(1.5, 12.0)$ \\
        $\beta$ & Spectral index for the PL of the mass ratio distribution. & $\mathcal{U}(-4.0, 12.0)$ \\
        $m_{\text{min}}$ & Minimum mass of the PL component of the primary mass distribution. & $\mathcal{U}(2.0\,M_{\odot}, 10.0\,M_{\odot})$ \\
        $m_{\text{max}}$ & Maximum mass of the PL component of the primary mass distribution. & $\mathcal{U}(50.0\,M_{\odot}, 200.0\,M_{\odot})$ \\
        \hline
        \multicolumn{3}{c}{\textsc{Power law + peak}} \\
        \hline
        \textbf{Parameter} & \textbf{Description} & \textbf{Prior} \\
        \hline
        $\alpha$ & Spectral index for the PL of the primary mass distribution. & $\mathcal{U}(1.5, 12.0)$ \\
        $\beta$ & Spectral index for the PL of the mass ratio distribution. & $\mathcal{U}(-4.0, 12.0)$ \\
        $m_{\text{min}}$ & Minimum mass of the PL component of the primary mass distribution. & $\mathcal{U}(2.0\,M_{\odot}, 10.0\,M_{\odot})$ \\
        $m_{\text{max}}$ & Maximum mass of the PL component of the primary mass distribution. & $\mathcal{U}(50.0\,M_{\odot}, 200.0\,M_{\odot})$ \\
        $\lambda_g$ & Fraction of the model in the Gaussian component. & $\mathcal{U}(0.0, 1.0)$ \\
        $\mu_g$ & Mean of the Gaussian component in the primary mass distribution. & $\mathcal{U}(20.0\,M_{\odot}, 50.0\,M_{\odot})$ \\
        $\sigma_g$ & Width of the Gaussian component in the primary mass distribution. & $\mathcal{U}(0.4\,M_{\odot}, 10.0\,M_{\odot})$ \\
        $\delta_m$ & Range of mass tapering at the lower end of the mass distribution. & $\mathcal{U}(0.0\,M_{\odot}, 10.0\,M_{\odot})$ \\
        \hline
        \multicolumn{3}{c}{\textsc{Broken power law}} \\
        \hline
        \textbf{Parameter} & \textbf{Description} & \textbf{Prior} \\
        \hline
        $\alpha_1$ & PL slope of the primary mass distribution for masses below $m_{\text{break}}$. & $\mathcal{U}(1.5, 12.0)$ \\
        $\alpha_2$ & PL slope for the primary mass distribution for masses above $m_{\text{break}}$. & $\mathcal{U}(1.5, 12.0)$ \\
        $\beta$ & Spectral index for the PL of the mass ratio distribution. & $\mathcal{U}(-4.0, 12.0)$ \\
        $m_{\text{min}}$ & Minimum mass of the PL component of the primary mass distribution. & $\mathcal{U}(2.0\,M_{\odot}, 10.0\,M_{\odot})$ \\
        $m_{\text{max}}$ & Maximum mass of the primary mass distribution. & $\mathcal{U}(50.0\,M_{\odot}, 200.0\,M_{\odot})$ \\
        $b$ & The fraction of the way between $m_{\text{min}}$ and $m_{\text{max}}$ at which the primary & $\mathcal{U}(0.0, 1.0)$ \\
        & mass distribution breaks. & \\
        $\delta_m$ & Range of mass tapering on the lower end of the mass distribution. & $\mathcal{U}(0.0\,M_{\odot}, 10.0\,M_{\odot})$ \\
        \toprule
    \end{tabular}
\end{table*}

To estimate the cosmological parameters and the source population properties of BBHs, we adopt the hierarchical Bayesian inference and employ the Python package \texttt{ICAROGW} \citep{2023arXiv230517973M}. 
Given the data $\{\mathbf{x}\}=\{x_{1},...,x_{\mathrm{\mathrm{obs}}}\}$ from $\mathrm{N_{\mathrm{obs}}}$ observations, we model the total number of events as an inhomogeneous Poisson process, yielding the likelihood function of the data given population parameters $\Phi$ and the intrinsic parameters $\theta$ (here the $\theta$ corresponds to $\theta_{D}$) as \citep{2019MNRAS.486.1086M,2023arXiv230517973M}
\begin{equation}
\label{eq:likelihood}
\mathcal{L}(x | \Phi) \propto \prod_{i=1}^{N_{\text{obs}}} \frac{\int \mathcal{L}\left(x_i| \theta, \Phi\right) \frac{\mathrm{d} N}{\mathrm{~d}t_{d}\mathrm{~d}\theta}\mathrm{d} \theta}{\int p_{\mathrm{det}}(\theta, \Phi) \frac{\mathrm{d} N}{\mathrm{~d} t_{d}\mathrm{~d}\theta} \mathrm{d} \theta}.
\end{equation}
Here, $\mathcal{L}(x_i | \theta, \Phi)$ is the individual likelihood for $i$th GW event, which can be derived from the individual posterior $p(\theta | x_{i},\Phi)$ by reweighing with the prior on intrinsic parameters $\theta$, where $p(\theta | x_{i},\Phi) \propto \mathcal{L}(x_{i}| \theta,\Phi) \pi(\theta | \Phi)$.
The conditional prior $\pi(\theta | \Phi)$ governs the population distribution on $\theta$ given a set of hyper-parameters $\Phi$ to describe the model. 
The detection probability $p_{\mathrm{det}}(\theta, \Phi)$ depends on $\theta$ that can be calculated as:
\begin{equation}
\label{eq:pdet}
    p_{\mathrm{det}}(\theta, \Phi)=\int_{x \in \mathrm{detectable}} \mathcal{L}(x_{i}| \theta, \Phi) \mathrm{d} x.
\end{equation}
The denominator of Eq.~\ref{eq:likelihood} represents the detection fraction, which quantifies the selection bias for a given set of population hyper-parameters $\Phi$. In general, Monte Carlo simulations of injected and detected events can be used to evaluate selection biases \citep{2021ApJ...913L...7A}, since we are unable to access a mathematical form of the detection probability \citep{2023AJ....166...22G}. In practice, it takes a set of detected injections $N_{\mathrm{det}}$ from the total injections $N_{\mathrm{gen}}$ generated from a prior distribution $\pi_{\mathrm{inj}}(\theta)$ to compute the integral using Monte Carlo integration, the specific formula for evaluating selection biases is given by Eq.~7 in \citet{2023arXiv230517973M}.



\subsection{Type Ia Supernovae Data}

We adopt the SNe Ia Pantheon+ compilation \citep{pantheonp}, which contains 1701 light curves of 1550 spectroscopically confirmed SNe Ia spanning the redshift range $0.00122<z<2.26137$.
This compilation is an extension of the Pantheon compilation \citep{pantheon}, with enhancements in sample size and improvements in the treatment of systematic uncertainties related to redshifts, peculiar velocities, photometric calibrations, and intrinsic scatter models of SNe Ia.
Due to the significant sensitivity of peculiar velocities at low redshifts $z < 0.008$, as displayed in Fig.~4 of Ref.~\citet{pantheonp}, which may lead to biased results, we exclude data points within the redshift range $z < 0.01$ in this work.

The observed distance modulus is defined as
\begin{equation}
\mu=m_{B}+\alpha x_{1}-\beta c-M-\delta_{\mathrm{bias}}+\delta_{\mathrm{host}},
\end{equation}
where $m_{B}$ denotes the apparent B-band magnitude, $x_{1}$ is the stretch parameter corresponding to light-curve width, $c$ is the light-curve color that includes contributions from both intrinsic color and dust, $M$ is the absolute B-band magnitude of a fiducial SNe Ia, $\delta_{\mathrm{bias}}$ is a correction term to account for selection biases, and $\delta_{\mathrm{host}}$ is the luminosity correction for residual correlations. $\alpha$ and $\beta$ are global nuisance parameters that relate stretch and color to luminosity, respectively. 

The theoretical distance modulus $\mu_{\mathrm{th}}$ is computed by 
\begin{equation}
    \mu_{\mathrm{th}}= 5\mathrm{log}_{10}\frac{D_{L}(z)}{\mathrm{Mpc}}+25,
\end{equation}
where $D_{\mathrm{L}}(z)$ is the luminosity distance associated with the cosmological parameters, given by
\begin{equation}
\label{eq:DL}
   D_L(z)=\frac{(1+z)}{H_0}\int^{z}_{0} \frac{d \tilde{z}}{E(\tilde{z})}.
\end{equation}

We follow the formalism outlined in \citet{pantheonp}, where cosmological parameters are constrained by minimizing a $\chi^{2}$ likelihood:
\begin{equation}
-2\mathrm{ln}(\mathcal{L})=\chi^{2}=\Delta \vec{D}^{T} \cdot \mathbf{C}^{-1} \cdot \Delta \vec{D},
\end{equation}
where the covariance matrix $\mathbf{C}$ is the covariance matrix including both the systematic and statistical errors, which can be found in the website\footnote{https://github.com/PantheonPlusSH0ES/DataRelease}, and $\Delta \vec{D}$ is the vector of SNe Ia distance modulus residuals,
\begin{equation}
    \Delta \vec{D}_{i}=\mu_{i}-\mu_{\mathrm{model}}(z_{i}).
\end{equation}

\subsection{Cosmological Models}
\label{sec:23}
In this work, we consider two typical cosmological models, the flat $\Lambda$CDM model and the flat $w$CDM model. We choose these models for two reasons. Firstly, they have long been famous for their success in describing the expansion history of the Universe. Secondly, given the current observations of GW and SNe Ia, it is challenging to effectively constrain complex cosmological models. 

For the flat $\Lambda$CDM model, the dimensionless Hubble parameter can be written as
\begin{equation}
H(z)=H_0 \sqrt{\Omega_{\mathrm{m}}(1+z)^{3}+\left(1-\Omega_{\mathrm{m}}\right)},
\end{equation}
where the cosmological parameters in $\Lambda$CDM model is $\Phi_{c}=(H_0,\Omega_{m})$.

For the flat $w$CDM model, the dimensionless Hubble parameter can be written as
\begin{equation}
H(z)=H_0 \sqrt{\Omega_{\mathrm{m}}(1+z)^{3}+(1-\Omega_{\mathrm{m}})(1+z)^{3(1+w)}},
\end{equation}
where $w=p/\rho$ is the EOS of dark energy, and the cosmological parameters in the $w$CDM model is $\Phi_{c}=(H_0,\Omega_{m},w)$. 
When $w=-1$, the $w$CDM model reduces to the $\Lambda$CDM model.

\subsection{Statistical Analysis Methods}
In previous studies \citep{LIGOScientific:2021aug}, the Bayes factors were calculated for comparing different mass models. Following this approach, we also adopt the Bayes factor \citep{bayes01,bayes02} to identify which mass model is more supported by the observations in this work. Furthermore, we employ the Bayesian Information Criterion (BIC) \citep{1978AnSta...6..461S} for additional analysis and discussion.

In the framework of Bayes' theorem, the posterior probability of the hypothesis $M_{i}$ considering the data $D$ can be expressed as
\begin{equation}
    \label{eq:bayesp}
    P(M_{i}| D)=\frac{P(D| M_{i})P(M_{i})}{P(D)},
\end{equation}
where $P(M_{i})$ represents the prior probability for the hypothesis $M_{i}$, $P(D | M_{i})$ denotes the likelihood function, for which we often use the shorthand notation $\mathcal{L}=P(D | M_{i})$, and $P(D)$ stands for the marginal likelihood (commonly referred to in cosmology as the ``Bayes evidence"), which can be written as
\begin{equation}
    P(D| M_{i})= \int P(D| \bar{\theta},M_{i})P(\bar{\theta}| M_{i})d\bar{\theta},
\end{equation}
where a set of parameters $\theta$ represents a specific model, $P(D| \bar{\theta},M_{i})$ is the likelihood function under the hypothesis $M_{i}$, and $P(\bar{\theta}| M_{i})$ is the prior probability for parameters $\bar{\theta}$ under the hypothesis $M_{i}$. 
Computing the Bayesian evidence requires the evaluation of an integral over the entire likelihood function and the prior of model parameters. 
When comparing two models, i.e., $M_{i}$ versus $M_{j}$, the Bayes factor $B_{i,j}$ is expressed as 
\begin{equation}
    B_{i,j}=\frac{P(D| M_{i})}{P(D| M_{j})}.
\end{equation}
When $\mathcal{B}=B_{i,j}>1$ (i.e., $\mathrm{log}_{10}\mathcal{B}>$0), it indicates that the observational data prefers model $M_{i}$ to model $M_{j}$ \citep{bayes03}.

The BIC can be computed as
\begin{equation}
\mathrm{BIC}=-2 \mathrm{ln} \mathcal{L}_\mathrm{max}+\mathrm{k} \mathrm{ln} \mathrm{N},
\end{equation}
where $\mathcal{L}_\mathrm{max}$ is the maximum likelihood of the model under consideration, $k$ is the number of model parameters, and $N$ is the number of data points. In general, the preferred model corresponds to the one that minimizes the BIC. For this reason, the weight of the evidence can be characterized by $\Delta \mathrm{BIC}=\mathrm{BIC}_{i}-\mathrm{BIC}_{\mathrm{min}}$, where the subindex $i$ refers to value of BIC for the model $i$ and $\mathrm{BIC}_{\mathrm{min}}$ is the minimum value of BIC among all the models.
It should be noted that $\Delta$BIC=2 indicates a positive evidence, while 2 $<\Delta$ BIC $<$ 6 represents a strong evidence against the model with a higher BIC value \citep{2004MNRAS.351L..49L}. 

Although these two methods can be used to compare models, the Bayes factor requires more computational resources as it directly evaluates the relative evidence between competing models within the Bayesian framework. In contrast, BIC offers a simpler criterion by penalizing model complexity, with the goal of identifying the model that is more favored by the observational data.

\section{Results and Discussion}
\label{sec:3}

\begin{table*}
\centering
    \caption{Values of the Hubble constant $H_0$, the matter density $\Omega_{m}$ and the equation of state of dark energy $w$ obtained in this study using 42 BBH events, SN, GW(42 BBH events and 1 BNS event) and GW(42 BBH events and 1 BNS event)+SN datasets, respectively, at the 68\% confidence levels (CLs). In the last three columns, we present the logarithm of the Bayes factor and the values of BIC between the different population mass models for the cases of $\Lambda$CDM and $w$CDM cosmology with different observational datasets}
    \label{tab:para}
    \begin{tabular}{llllllll}
    \hline
    \hline
    \multicolumn{8}{c}{\textbf{$\Lambda$CDM Model}} \\
    \hline
    \textbf{Data} & \textbf{Mass Model} & $\Omega_{m}$ & $H_{0}$(km/s/Mpc) &  & $\mathrm{BIC}$ & $\Delta$BIC  & $\mathrm{log}_{10}\mathcal{B}$ \\
    \hline
    42 BBH events & \textsc{Truncated} & $0.51\pm0.32$ & $51.90^{+77.89}_{-5.74}$ & \\
     & \textsc{Power law + peak} & $0.47\pm0.32$ & $64.13^{+25.98}_{-23.80}$ & \\
     & \textsc{Broken power law} &  $0.47\pm 0.32$ & $47.42^{+51.82}_{-7.10}$ & \\
    \hline
     SN & - & $0.30\pm0.02$ & - & \\
    \hline
    GW & \textsc{Truncated} & - & $71.53^{+27.55}_{-9.41}$ &  & 2298.21 &  +25.52 & -1.81  \\
     & \textsc{Power law + peak} & - & $70.97^{+12.50}_{-9.48}$ & &2272.69 & 0 & 0\\
     & \textsc{Broken power law} & - & $71.71^{+16.71}_{-10.69}$ & & 2293.23 & +20.54 & -0.29\\
     \hline
     GW+SN & \textsc{Truncated} & $0.36\pm0.03$ & $70.04^{+31.42}_{-8.33}$ & &  2799.58 & +22.05 & -2.59  \\
     & \textsc{Power law + peak} & $0.36\pm0.03$ & $69.44^{+14.84}_{-7.53}$ &  & 2777.53 & 0 &   0\\
     & \textsc{Broken power law} & $0.36\pm0.03$ & $70.31^{+17.93}_{-9.18}$ & & 2797.22 &+19.69 & -0.37\\
     \hline
     \hline
    \multicolumn{8}{c}{\textbf{$w$CDM Model}} \\
    \hline
    \textbf{Data} & \textbf{Mass Model} & $\Omega_{m}$ & $H_{0}$(km/s/Mpc) & $w$ & $\mathrm{BIC}$ & $\Delta$BIC  & $\mathrm{log}_{10} \mathcal{B}$\\
    \hline
    GW & \textsc{Truncated} & - & $73.71^{+21.47}_{-11.59}$ & - & 2301.81 & +27.95 &   -2.22  \\
    & \textsc{Power law + peak} & - & $69.97^{+15.88}_{-8.40}$ & - & 2273.86 & 0 & -0.57 \\
    & \textsc{Broken power law} & - & $70.48^{+15.17}_{-8.68}$ & - & 2296.70 & +22.84 &  0\\
    \hline
    GW+SN & \textsc{Truncated} & $0.27^{+0.14}_{-0.11}$ & $69.29^{+31.51}_{-8.45}$ & $-0.86^{+0.30}_{-0.14}$ & 2814.00 &+29.75   &  -2.02 \\
    & \textsc{Power law + peak} &  $0.27^{+0.13}_{-0.11}$ & $ 70.91^{+12.65}_{-9.43}$ & $-0.84^{+0.27}_{-0.13}$ & 2784.25 & 0   & 0\\
    & \textsc{Broken power law} & $0.25^{+0.13}_{-0.12}$ & $72.09^{+12.65}_{-11.57}$ & $-0.82^{+0.27}_{-0.12}$ & 2803.73 &  +19.48 &  -0.23  \\
    \hline
    \hline
\end{tabular}
\end{table*}

\begin{figure*}
\centering
\subfigure[\textsc{Truncated}]{\includegraphics[width=0.3\linewidth]{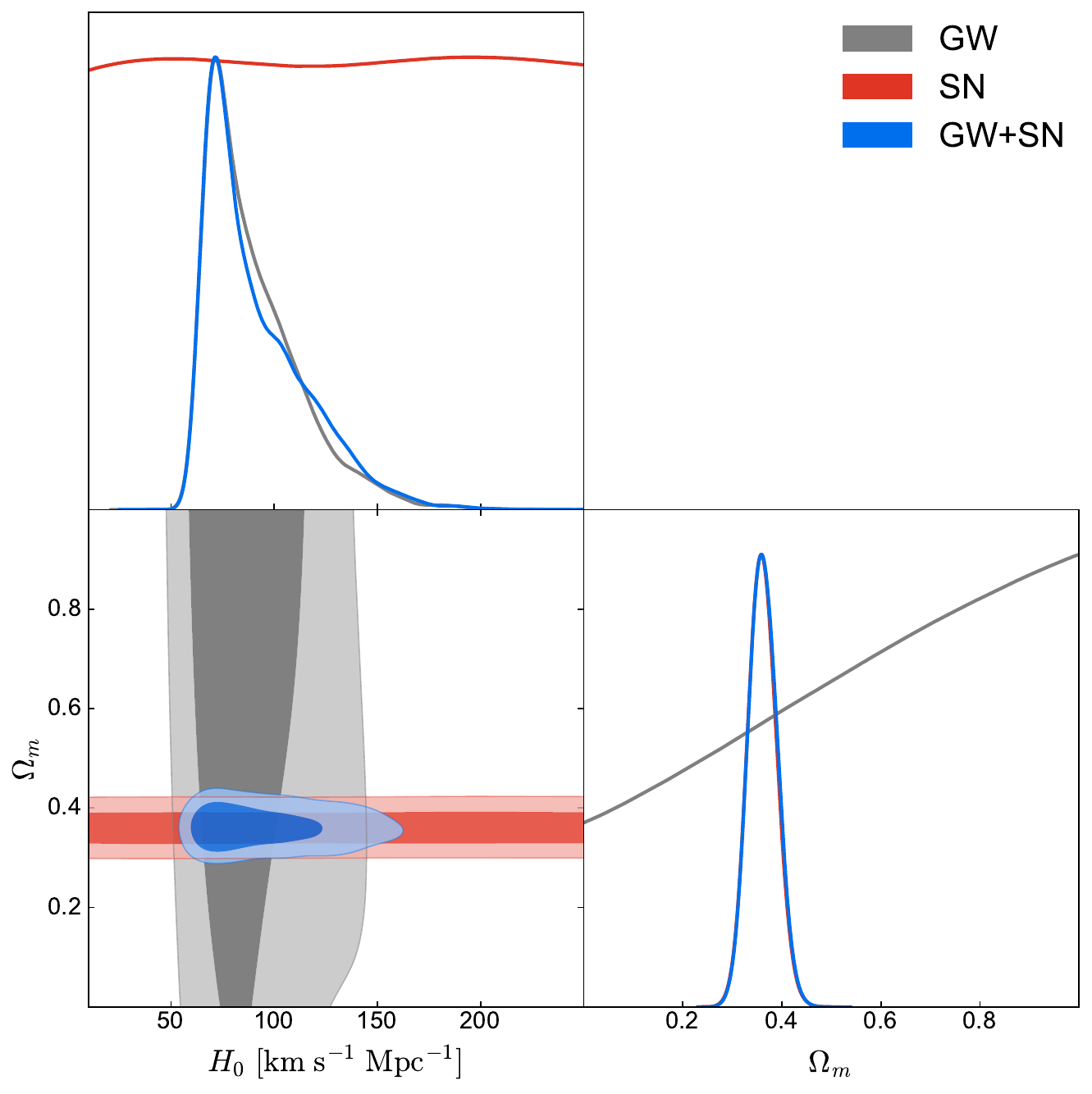}}
\subfigure[\textsc{Power law + peak}]{\includegraphics[width=0.3\linewidth]{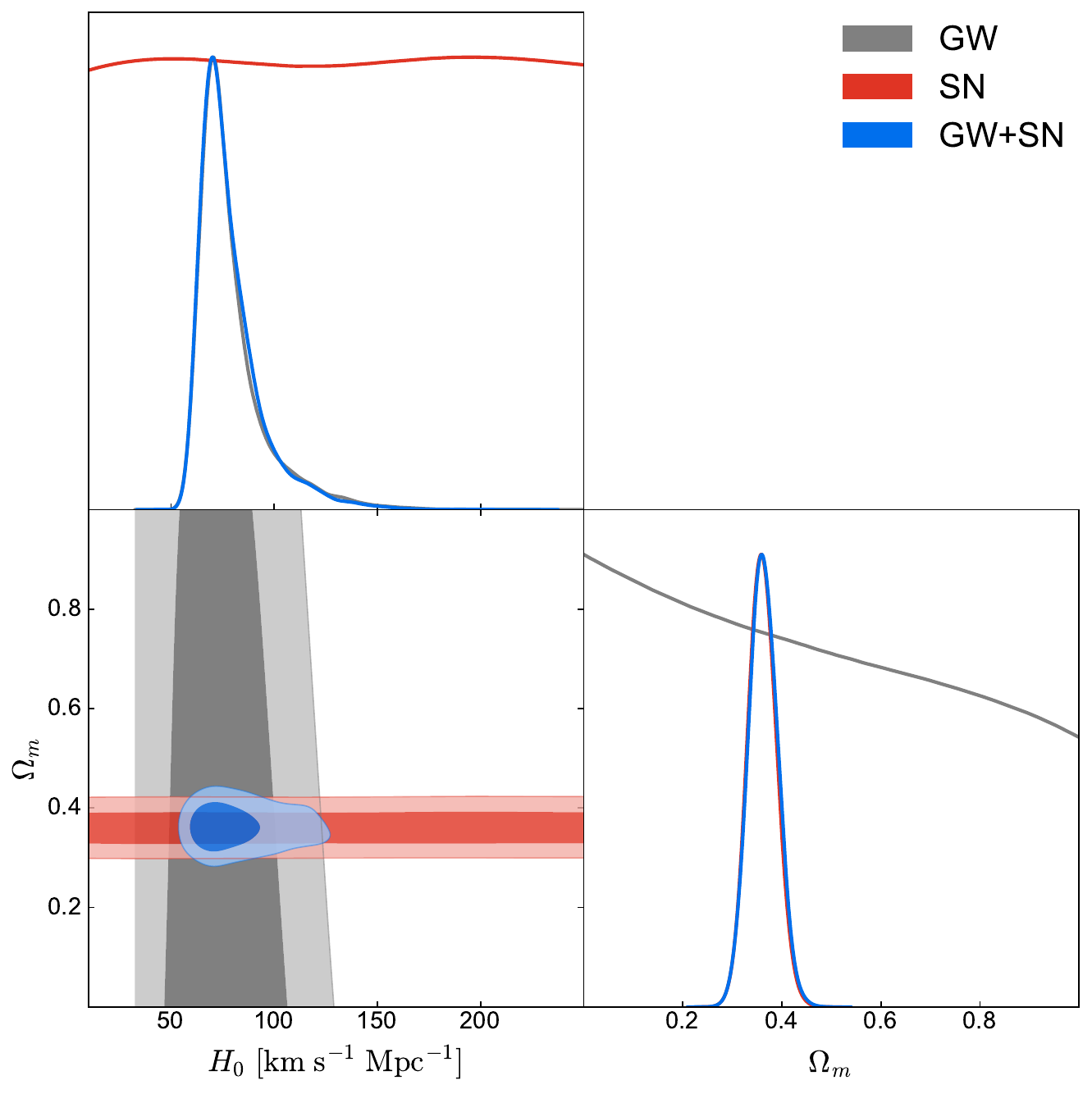}}
\subfigure[\textsc{Broken power law}]{\includegraphics[width=0.3\linewidth]{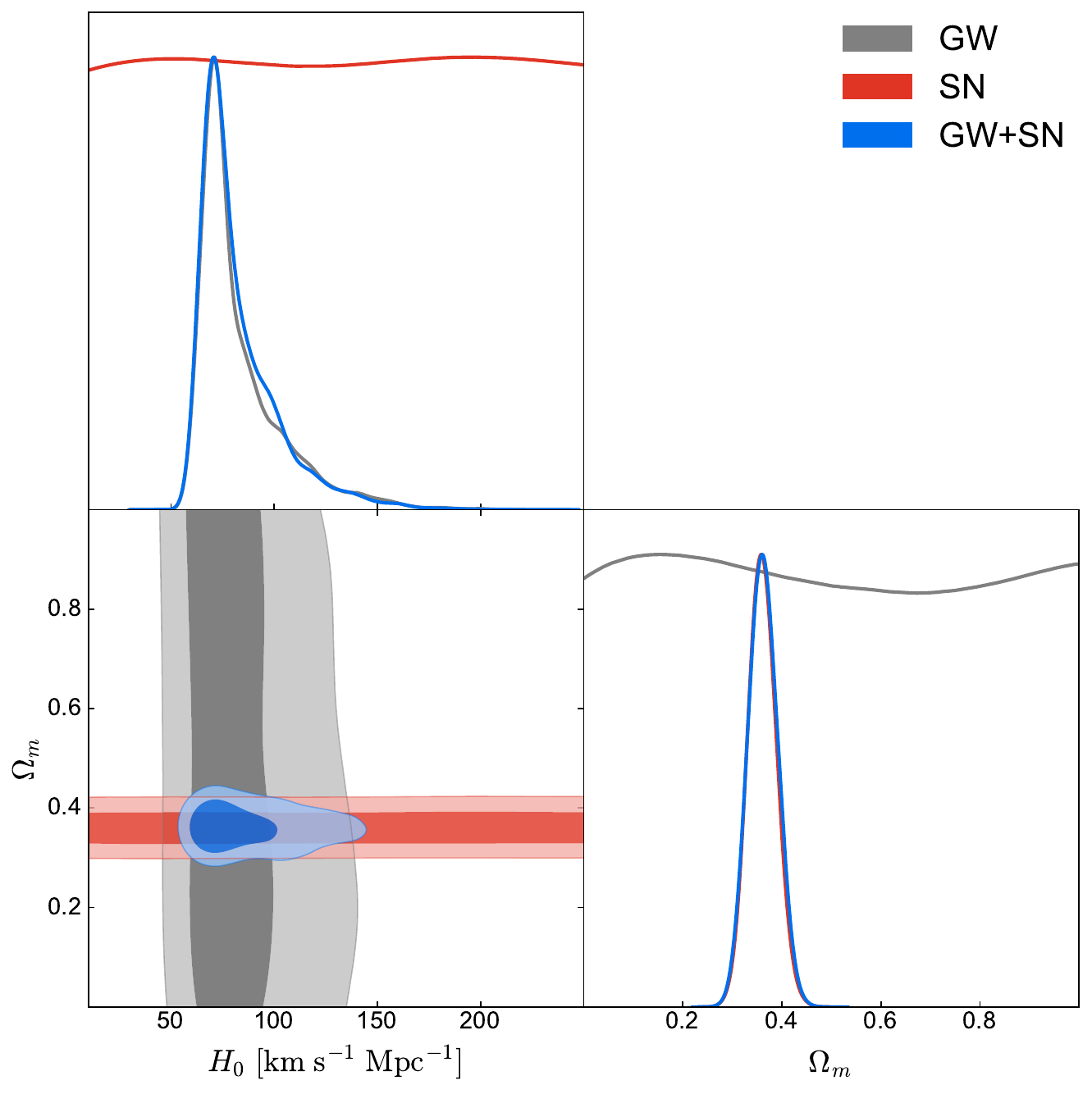}}
\caption{In each panel, the posterior probability density is depicted for the $\Lambda$CDM model within three phenomenological population mass models. The solid lines delineate the 68\% and 95\% confidence level (CL) contours, respectively. The gray lines represent the constraints from GW data, the red lines represent the SNe Ia data, and the blue lines represent the combination of SNe Ia and GW.}
\label{fig:lcdm}
\end{figure*}

\begin{figure*}
\centering
\subfigure[\textsc{Truncated}]{\includegraphics[width=0.3\linewidth]{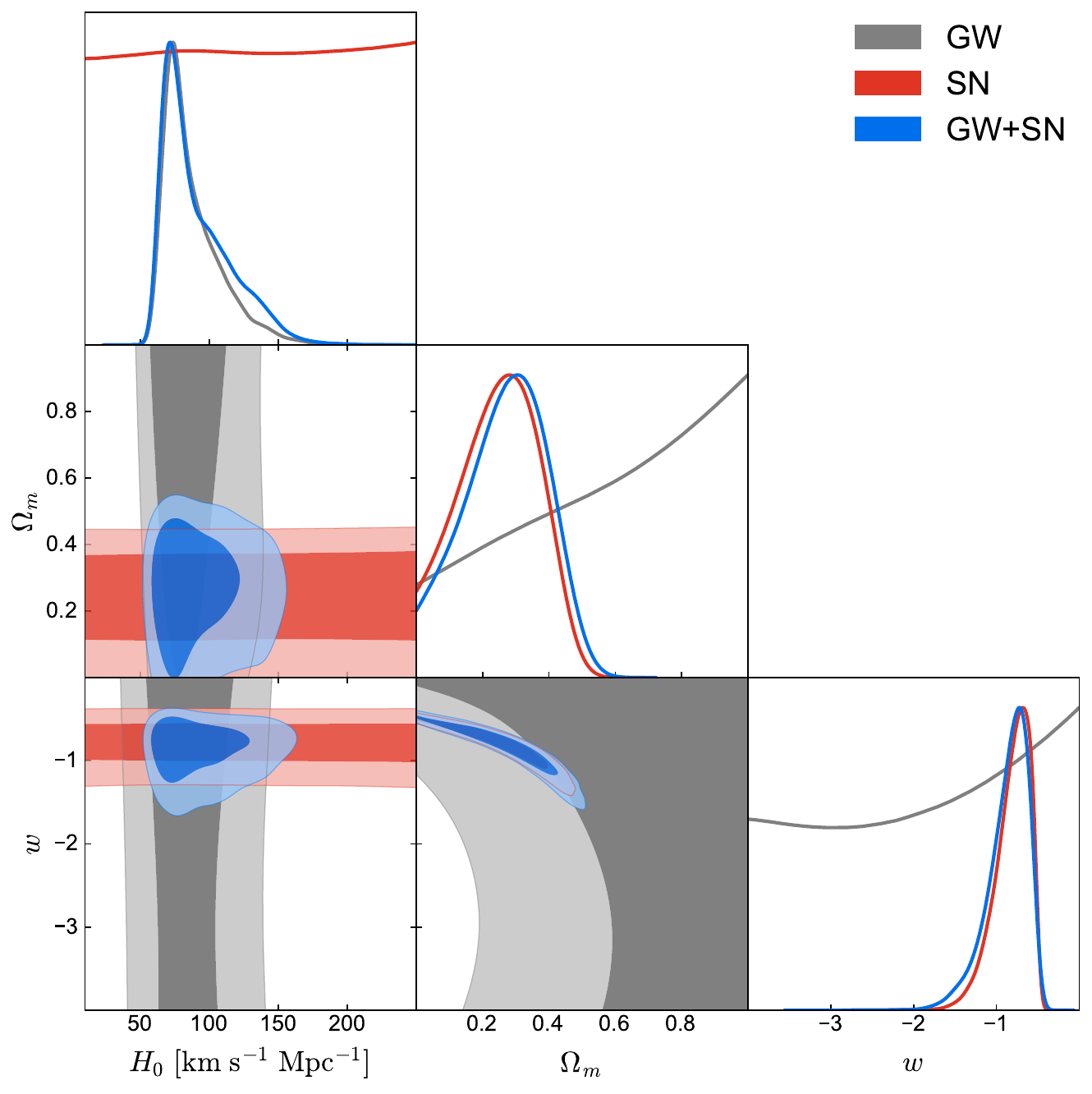}}
\subfigure[\textsc{Power law + peak}]{\includegraphics[width=0.3\linewidth]{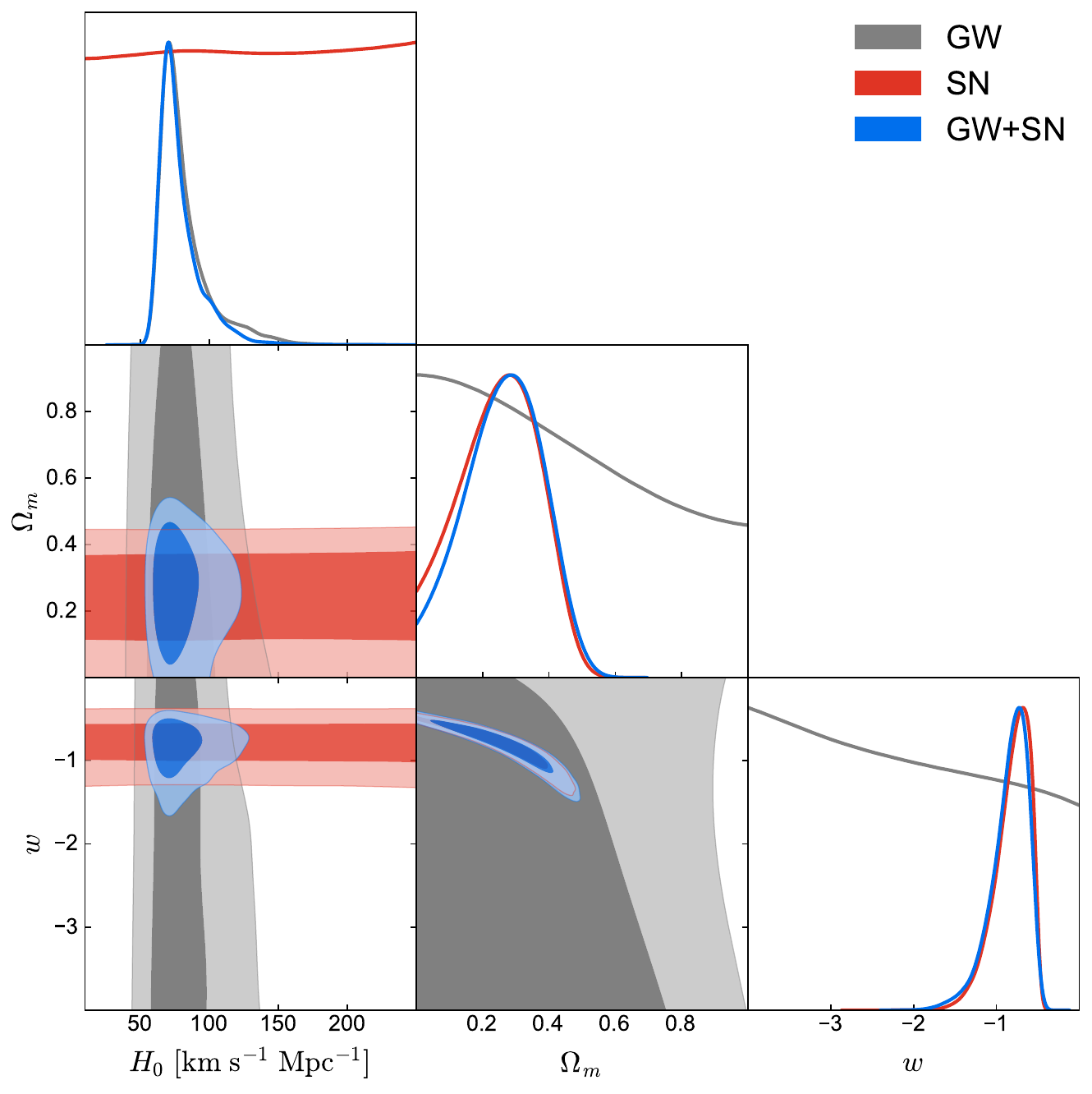}}
\subfigure[\textsc{Broken power law}]{\includegraphics[width=0.3\linewidth]{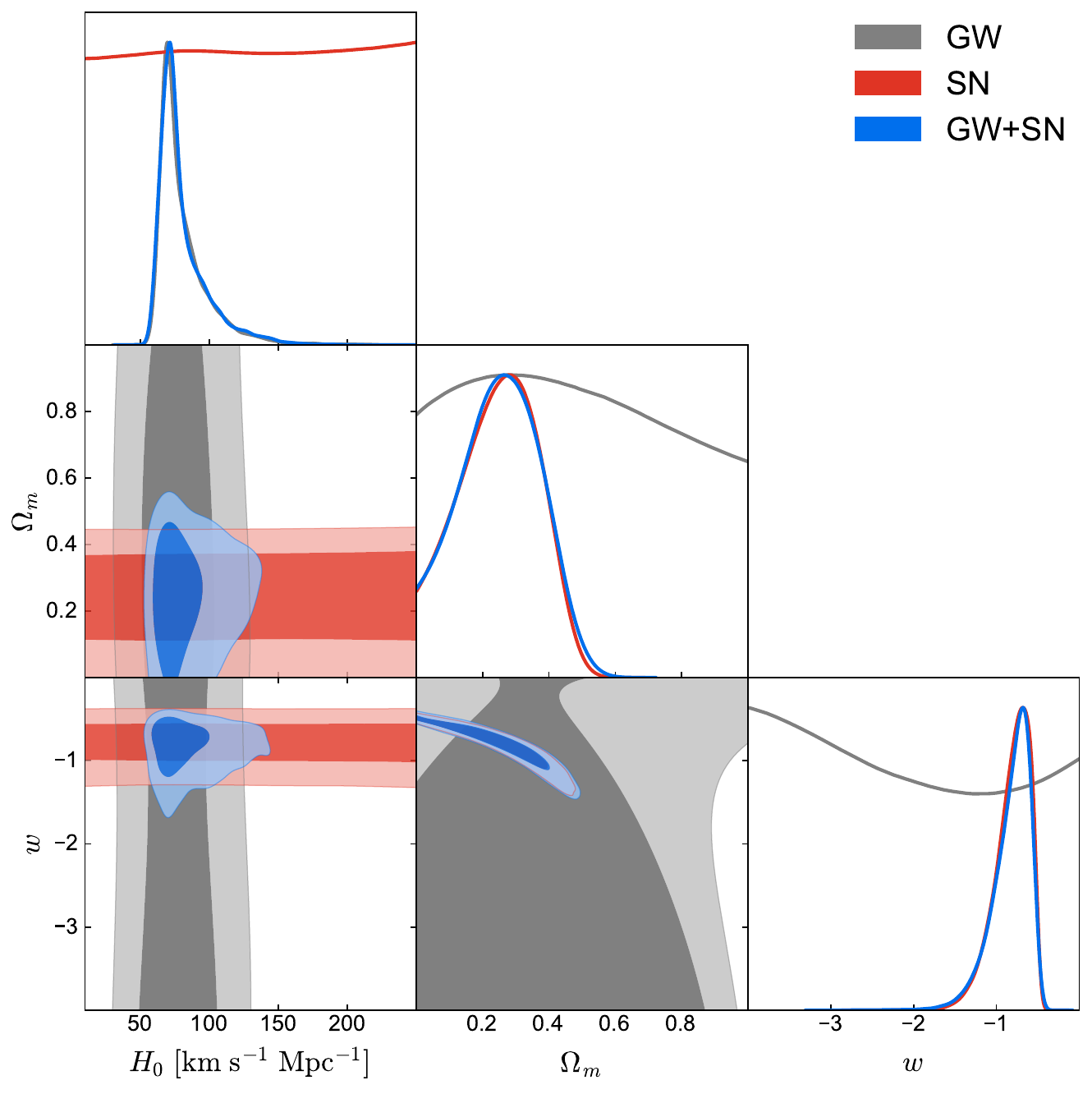}}
\caption{In each panel, the posterior probability density is depicted for the $w$CDM model within three phenomenological population mass models. The solid lines delineate the 68\% and 95\% confidence level (CL) contours, respectively. The gray lines represent the constraints from GW data, the red lines represent the SNe Ia data, and the blue lines represent the combination of SNe Ia and GW.}
\label{fig:wcdm}
\end{figure*}

GW observations provide absolute distance measurements, while SNe Ia observations offer relative distances. In this section, we present cosmological constraints derived from these two probes to investigate their ability to break degeneracies between cosmological parameters. The cosmological constraints are implemented using the \texttt{emcee} \citep{emcee} and \texttt{ICAROGW} \citep{2021PhRvD.104f2009M,2023arXiv230517973M} Python packages.

Tab.\ref{tab:para} lists the best-fit values of the cosmological parameters within 1$\sigma$ confidence level (CL), and the Bayes factor and BIC values for each phenomenological mass model.
Fig.~\ref{fig:lcdm} illustrates the posterior distributions of $H_0$ and $\Omega_{m}$ for the $\Lambda$CDM model. GW data alone yield $H_{0}=71.53^{+27.55}_{-9.41}$ km/s/Mpc, $70.97^{+12.50}_{-9.48}$ km/s/Mpc, and $71.71^{+16.71}_{-10.69}$ km/s/Mpc for the \textsc{Truncated}, \textsc{Power law + peak}, and \textsc{Broken power law} models with 1$\sigma$ CLs, respectively, with the \textsc{Power law + peak} model offering the highest precision of 17.61\% for $H_0$. However, GW data alone do not provide tight constraints on $\Omega_{m}$. In contrast, SNe Ia data constrain $\Omega_{m}$ to $0.30\pm0.02$ with 1$\sigma$ CL, with a precision of 8\%.

The combination of GW and SNe Ia data yields $H_{0}=70.07^{+31.42}_{-8.33}$ km/s/Mpc, $69.44^{+14.84}_{-7.53}$ km/s/Mpc, and $70.31^{+17.93}_{-9.18}$ km/s/Mpc for the \textsc{Truncated}, \textsc{Power law + peak}, and \textsc{Broken power law} models at 1$\sigma$ CLs, respectively. Although the combined GW and SNe Ia data yield the highest precision of 21.73\% on $H_0$ and 8.3\% on $\Omega{m}$ for the \textsc{Power law + peak} model, it is important to note that high precision in parameter estimation does not mean that this population mass model is in good agreement with observations. Further discussion on the phenomenological mass models will be presented in the following content.

The orthogonal orientations of the $H_{0}-\Omega_{m}$ planes, as seen in the grey and red contours in Fig.~\ref{fig:lcdm}, demonstrate that the combination of GW and SNe Ia data can effectively break parameter degeneracies.
These results indicate the advantages of combining GW and SNe Ia data for both constraining cosmological parameters and breaking parameter degeneracies. Nevertheless, the current level of precision is insufficient to meet the requirements of precision cosmology, highlighting the need for further improvements in observational capabilities and analysis techniques.

Fig.~\ref{fig:wcdm} shows the posterior distributions on $H_0$, $\Omega_{m}$ and $w$ for the $w$CDM model.
When considering the GW data alone, we find $H_{0}=73.71^{+21.47}_{-11.59}$ km/s/Mpc for the \textsc{Truncated} model, $H_{0}=69.97^{+15.88}_{-8.40}$ km/s/Mpc for the \textsc{Power law + peak} model, and $H_{0}=70.48^{+15.17}_{-8.68}$ km/s/Mpc for the \textsc{Broken power law} model, at 1$\sigma$ CLs, respectively.
When combining the GW and SNe Ia data, we obtain $H_{0}=69.29^{+31.51}_{-8.45}$ km/s/Mpc, $\Omega_{m}=0.27^{+0.14}_{-0.11}$ and $w=-0.86^{+0.30}_{-0.14}$ for the \textsc{Truncated} model, $H_{0}=70.91^{+12.65}_{-9.43}$ km/s/Mpc, $\Omega_{m}=0.27^{+0.13}_{-0.11}$ and $w=-0.84^{+0.27}_{-0.13}$ for the \textsc{Power law + peak} model, and $H_{0}=72.09^{+12.65}_{-11.57}$ km/s/Mpc, $\Omega_{m}=0.25^{+0.13}_{-0.12}$ and $w=-0.82^{+0.27}_{-0.12}$ for the \textsc{Broken power law} model, at 1$\sigma$ CLs, respectively.
It shows that the \textsc{Broken power law} model provides the highest precision on cosmological parameters. The GW data alone can provide a precision of 21.52\% on $H_0$, whereas the combination of GW and SNe Ia data provides a precision of 17.55\% on $H_0$, a precision of 52\% on $\Omega_{m}$ and a precision of 32.93\% on $w$. 
It has been demonstrated that in the framework of the $w$CDM model, the combination of GW and SNe Ia data fails to effectively constrain the parameters $\Omega_{m}$ and $w$. This may due to the fact that the $w$CDM model is more complex than the standard $\Lambda$CDM model, proving that the current GW and SNe Ia observations cannot constrain complex model cosmological models.
In previous studies, the $w$CDM model has been investigated using the simulated GW data generated by the Einstein Telescope (ET) and the Cosmic Explorer (CE), which reveals that even with significant improvements in both accuracy and volume of GW data, the constraints on $w$ remain weak \citep{2017PhRvD..95d4024C,2019JCAP...09..068Z,2020EPJC...80..217Z,zheng2022}.
Moreover, our constraint results based on the \textsc{Power law + peak} and the \textsc{Broken power law} mass models show 1$\sigma$ deviation from the $\Lambda$CDM model. Considering that the current accuracy cannot meet the requirement of precision cosmology, it is essential to combine these two datasets with other cosmological probes that may help to enhance the accuracy of $w$.
Based on the constraint results of $H_0$ we obtained, it is evident that they are broadly consistent with the recent $H_0$ estimations \citep{planck2018,riess}. However, these results do not contribute to explore the Hubble tension problem, as the errors on $H_0$ estimation extend beyond the Hubble tension region.

\begin{figure*}
\centering
\subfigure[]{\includegraphics[width=2.5in]{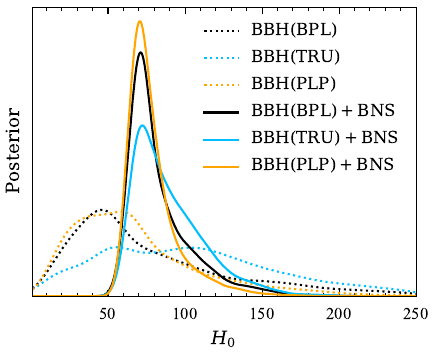}}
\subfigure[]{\includegraphics[width=2.5in]{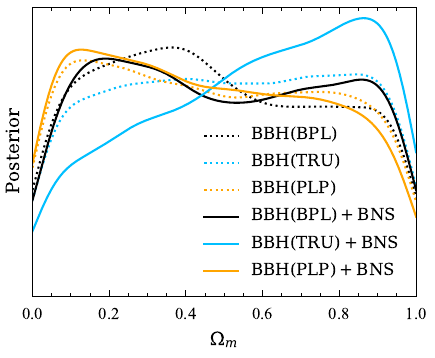}}
\caption{Posterior distributions for $H_{0}$ and $\Omega_{m}$ obtained from the 42 BBH events (dotted lines) and combined with the bright standard siren GW170817 (solid lines)}
\label{fig:lcdm-1d}
\end{figure*}

Considering that we do not yet have a mass model for black holes and neutron stars, we exclude two NSBH events, GW200105 and GW200115, the BNS event GW190425, and the dubious event GW190814 from our analysis. Hence, in Fig.~\ref{fig:lcdm-1d}, we compare the posterior distributions for $H_0$ and $\Omega_{m}$ derived from 42 BBH events alone with the combined $H_0$ and $\Omega_{m}$ posteriors from GW170817, respectively.
We find that the inclusion of BNS event significantly enhances the precision of $H_0$. 
This suggests that observations of EM counterparts will play a crucial role in constraining $H_0$. As we discussed before, the population assumptions can indeed affect the constraint results, but when combined with the BNS event GW170817, the population assumptions have less impact on the estimation of $H_0$.
The current precision of $H_0$ from GW observations remains limited, the future enhancement of detection sensitivity and the increasing number of detections would help significantly improve the constraint on $H_0$ \citep{2023JCAP...08..070J,2023arXiv230911900J,2024SCPMA..6730411S}.
On the other hand, whether bright siren or dark siren, the constraints on $\Omega_{m}$ are relatively weak.


In Tab.~\ref{tab:para}, we find that the \textsc{Truncated} model is strongly disfavored compared to the \textsc{Power law + peak} and \textsc{Broken power law} models, with a $\mathrm{log}_{10}\mathcal{B}\sim -2$.
This relatively poor fit is due to the inability of the \textsc{Truncated} model to accommodate the high-mass BBH events \citep{BBH,2021ApJ...913L...7A,ligod,LIGOScientific:2021aug}.
In the case of the \textsc{Truncated} model, broader posteriors on cosmological parameters are observed, indicating that the knowledge of the source mass distribution would influence cosmological parameter estimation.
According to the values of $\mathrm{log}_{10}\mathcal{B}$, there is no compelling evidence to prefer the \textsc{Power law + peak} model over the \textsc{Broken power law} model.
As depicted in Tab.~\ref{tab:para}, the \textsc{Broken power law} model yields constraint results that are qualitatively similar to those obtained from the \textsc{Power law + peak} model.
Our results of the Bayes factors are consistent with Ref.~\citet{2021ApJ...913L...7A,ligod,2021PhRvD.104f2009M,LIGOScientific:2021aug}. 
In addition, we employ the BIC to determine the most favored mass model by the current observations.
It reveals that the \textsc{Power law + peak} model is strongly favored with respect to the \textsc{Truncated} model and the \textsc{Broken power law} model, as the \textsc{Power law + peak} model owns the lowest values of $\Delta \mathrm{BIC}$.
From the above results, it can be seen that the \textsc{Truncated} model is disfavored compared to the more complicated models, and the \textsc{Power law + peak} model owns modest support among the three mass models.
Remarkably, the \textsc{Power law + peak} model also exhibits the highest precision in constraining the parameters.

It is essential to compare the results derived from the population method with those from the galaxy catalog method.
In Ref.~\citet{2023ApJ...943...56P}, a $H_0$ value of $79.8^{+19.1}_{-12.8}$ km/s/Mpc with approximately 20\% precision was reported, derived from 8 well-localized dark sirens combined with the galaxy catalog method. Additionally, they obtained $H_0=72.77^{+11.0}_{-7.55}$ km/s/Mpc with roughly 12\% precision by combining the posterior with that from GW170817.
In Ref.~\citet{2023JCAP...12..023G}, a reanalysis of GWTC-3 with the GLADE+ galaxy catalog yielded $H_0=69^{+12}_{-7}$ km/s/Mpc with a precision of approximately 20\%.
Furthermore, in Ref.~\citet{LIGOScientific:2021aug}, they used 47 GW events from GWTC-3 with the GLADE+ galaxy catalog, resulting in a measurement of $H_{0}$ with roughly 19\% precision.
In our research, it is clear that we not only derive constraint results of $H_0$ that are consistent with those obtained from the galaxy catalog method, but also effectively and simultaneously constraints on $H_0$, $\Omega_{m}$, and $w$.
The galaxy catalog method requires fixing the BH population when computing the selection function and the BH population assumption would strongly impact the out-of-catalog term of the likelihood. However, it is less important for events with a small localization in a region of the galaxy catalog that is complete, such as GW190814.
Moreover, it is worth noting that the GLADE+ galaxy catalog can be considered complete only up to $z \sim 0.011$, with completeness declining to 20\% at $z \sim 0.167$, limiting the capability of the galaxy catalog method \citep{2024SCPMA..6730411S}.
The upcoming galaxy surveys, such as the Dark Energy Spectroscopic Instrument (DESI) survey \citep{2016arXiv161100036D}, the Large Synoptic Survey Telescope (LSST) \citep{2009arXiv0912.0201L}, the Euclid space mission \citep{2011arXiv1110.3193L}, and the China Space Station Telescope (also known as the Chinese Survey Space Telescope, CSST) \citep{2018MNRAS.480.2178C,2022MNRAS.511.1830C}, hold great potential for precisely measuring cosmological parameters via the galaxy catalog method.




\section{Conclusion}
\label{sec:4}
Discrepancies in measurements of some significant cosmological parameters suggest that our understanding of the Universe may be inadequate, motivating us to re-examine fundamental questions using new and independent probes.
Combining the late-Universe probes offers an opportunity to break the degeneracies of parameters and enable to achieve the precision cosmology, which is beneficial for exploring the nature of dark energy and resolving the inconsistencies between the early and late Universe \citep{Lian:2021tca,Cao:2021zpf,Liu:2022mpj,zheng_quasar,zheng2022,Qi:2023oxv}. 
The abundance of observed GW events by ground-based detectors carries significant implications for cosmology. 
Considering the complementarity between GW and SNe Ia, their combination is expected to play a crucial role in cosmology.
In this paper, we represent an analysis for constraining the dark energy models with the 43 GW events from the GWTC-3 and 1590 SNe Ia data from the ``Pantheon+" sample. We have inferred constraints on the cosmological parameters of the $\Lambda$CDM and $w$CDM models.
For BBH events, we adopt the population method to infer the redshift information. Here, three population mass models are considered, the \textsc{Truncated} model, the \textsc{Power law + Peak} model and the \textsc{Broken power law} model. The main conclusions can be summarized as follows:

(i) GW data alone cannot tightly constrain cosmological parameters except for $H_{0}$, which can be an important supplement to the SNe Ia data. The results show that the parameter degeneracy directions from GW and SNe Ia data are rather different. Therefore, their combination can effectively break the degeneracies between parameters and give tighter constraints on cosmological parameters, which offer an alternative way to solve the inconsistencies between the early and late Universe. However, our results indicate that with the current GW and SNe Ia data, the constraints on cosmological parameters are still far from the standard of precision cosmology. 

(ii) In the $\Lambda$CDM model, by combining the GW data (42 BBH events and 1 BNS event) with the SNe Ia data, the uncertainties on $H_0$ and $\Omega_{m}$ are measured to be approximately 20\% and 8\% based on the \textsc{Power law + peak} mass model. In contrast, for the $w$CDM model, the uncertainties on $H_0$, $\Omega_{m}$ and $w$ are measured to be approximately 17\%, 50\% and 30\%, respectively, based on the \textsc{Broken power law} mass model. It is clear that the constraints on the Hubble constant $H_0$, the present-day matter density $\Omega_{m}$ and dark energy EoS, $w$, parameters are relatively weak, and they are still far from the standard of precision cosmology.


(iii) In our work, we present constraints on $w$CDM model, where the EoS of dark energy parameter $w$ is constant in time. This model reduces to $\Lambda$CDM when $w=-1$.
We find that $w=-0.84^{+0.27}_{-0.13}$ and $w=-0.82^{+0.27}_{-0.12}$ from the combination of GW and SNe Ia, based on the \textsc{Power law + peak} and \textsc{Broken power law} mass model, respectively, at 68\% CLs.
These results for the $w$CDM indicate an approximately 1$\sigma$ difference to the $\Lambda$CDM model.
Nevertheless, these findings provide a tantalizing suggestion of deviations from the standard cosmological model that motivate further research and highlight the potential of the combination of GW and SNe Ia observations to better understand the nature of dark energy.

(iv) According to the results of Bayes factor and BIC, it can be observed that the data strongly disfavor the \textsc{Truncated} mass model, which is consistent with the findings in \citet{LIGOScientific:2021aug}. Additionally, the \textsc{Power law + peak} and the \textsc{Broken power law} mass models are modestly preferred by the observations, as indicated by their smaller Bayes factor and BIC values. Furthermore, the results obtained from these models exhibit the highest precision of $\Lambda$CDM and $w$CDM models, respectively. However, we still cannot determine which of these two population mass models is most favorable.

(v) When compared to the BBH events, the BNS event shows an outstanding capability on constraining $H_0$. Therefore, the detection of EM counterparts plays a significant role in measuring $H_0$. However, with the increasing number of BBH events, the ability of dark sirens to constrain $H_0$ will be greatly enhanced.

In summary, combining various late-Universe probes offers a new way to resolve the fundamental issues in cosmology, such as the Hubble tension.
It is promising that the combination of future GW and SNe Ia observations will break the parameters degeneracy and play a crucial role in estimating cosmological parameters.

The next-generation GW survey is anticipated to increase the number of GW detections to several hundreds, with a BNS detection range increased by approximately 15\%-40\% \citep{2018LRR....21....3A,2023JCAP...12..023G}. The measurement of bright sirens will significantly aid in inferring the redshift from direct observations of EM counterparts. Moreover, the space-based GW observatories have also been put forward, such as the Laser Interferometer Space Antenna (LISA) \citep{2003PhRvD..67b9905C}, TianQin and Taiji projects \citep{2016gac..conf..360M,2018arXiv180709495R,2022PhRvD.105b2001L}. These space-based detectors aim to detect low-frequency GWs primarily generated from supermassive black hole coalescence and extreme mass ratio inspirals, providing a large sample of well-measured observations of BBH mergers. 
The combination of advanced gravitational wave detectors, including ground-based and space-based observatories, along with refined analysis methods, will enable a precise measurement of the Hubble constant.
Furthermore, numerous EM surveys are planned for construction and operation, including the upcoming datasets from SN programs, such as the Legacy Survey of Space and Time (LSST) \citep{2009arXiv0912.0201L}, Zwicky Transient Facility (ZTF) \citep{2022MNRAS.510.2228D}, Young Supernova Experiment (YSE) \citep{2021ApJ...908..143J}, etc.
Besides, the LSST is anticipated to process approximately $10^6$ transient detections per night, potentially increasing the SNe Ia sample size by up to a factor of 100 compared to previous samples \citep{2022ApJ...934...96S}. 
These future surveys are expected to provide more robust estimations of cosmological parameters, which can help to resolve inconsistencies between the early and late Universe.
Moreover, the observations of multi-messenger and multi-wavelength, including optical, infrared, radio, and GW, will enable precise measurements of cosmological parameters, paving the way to address fundamental questions in cosmology.

\section*{Acknowledgments}
This work was supported by the National SKA Program of China (Grants Nos. 2022SKA0110200 and 2022SKA0110203), and the National Natural Science Foundation of China (Grants Nos. 12205039). Jingzhao Qi is funded by the China Scholarship Council.

\appendix
\section{Population Mass Models}
As mentioned before, we use three phenomenological mass models in this paper.
The first model is the \textsc{Truncated} model, which characterizes the distribution of the primary mass $m_{1}$ using a truncated power law with a slope of $-\alpha$ between a minimum mass $m_{\mathrm{min} }$ and a maximum mass $m_{\mathrm{max}}$ \citep{2017ApJ...840L..24F}. This mass distribution $p(m_{1}|\Phi)=p\left(m_{1} | m_{\mathrm{min} }, m_{\mathrm{max} }, \alpha\right)$ takes the form as 
\begin{equation}
p\left(m_{1} | m_{\mathrm{min} }, m_{\mathrm{max} }, \alpha\right)=\mathcal{P}\left(m_{1} | m_{\min }, m_{\max },-\alpha\right),
\end{equation}
where the truncated power law distribution is described by slope $\alpha$, and lower and upper bounds $x_{\mathrm{min}}$, $x_{\mathrm{max}}$ at which there is a hard cut-off,
\begin{equation}
\label{eq:tpl}
\mathcal{P}(x | x_{\mathrm{min}}, x_{\mathrm{max}} , \alpha)= \begin{cases}x^{\alpha}, & (x_{\mathrm{min}}\leq x \leq x_{\mathrm{min}}) \\ 0, & \text { Otherwise. }\end{cases}
\end{equation}

The second model is the \textsc{Power law + peak} model, which consists of a truncated power-law with a slope of $-\alpha$ between a minimum mass $m_{\mathrm{min} }$ and a maximum mass $m_{\mathrm{max} }$, along with a Gaussian component distribution with a mean of $\mu_{\mathrm{g}}$ and a standard deviation of $\sigma_{g}$ \citep{2018ApJ...856..173T}, defined as 
\begin{equation}
    \mathcal{G}(x | \mu, \sigma, a, b)= \frac{1}{\sigma \sqrt{2 \pi}} \mathrm{exp} [-\frac{(x-\mu)^2}{2 \sigma^2}].
\end{equation}
And, this mass model can be expressed as 
\begin{equation}
\begin{split}
p\left(m_{1} | m_{\min }, m_{\max }, \alpha, \lambda_{\mathrm{g}}, \mu_{\mathrm{g}}, \sigma_{\mathrm{g}}\right)&= \\
[\left(1-\lambda_{\mathrm{g}}\right) \mathcal{P}\left(m_{1} | m_{\min }, m_{\max },-\alpha\right) & +\lambda_{\mathrm{g}} \mathcal{G}\left(m_{1} | \mu_{\mathrm{g}}, \sigma_{\mathrm{g}}\right)],
\end{split}
\end{equation}
where $\lambda_{\mathrm{g}}$ is a ratio parameters of these two component $\mathcal{P}$ and $\mathcal{G}$, the truncated power law distribution is the same as Eq.\ref{eq:tpl}.

The third model is the \textsc{Broken power law} model, which is characterized by a double truncated power-law distribution with slopes $\alpha_1$ and $\alpha_2$, and by a breaking point between the two regimes at $m_{\mathrm{break}}=b(m_{\mathrm{max}}-m_{\mathrm{min}})$ \citep{2018ApJ...856..173T,2021ApJ...913L...7A}, where $b\in [0,1]$,
\begin{equation}
\begin{split}
p(m_{1} | m_{\min }, m_{\max }, \alpha_{1}, \alpha_{2})=[\mathcal{P}(m_{1} | m_{\min }, m_{\mathrm{break}},-\alpha_1)\\
+\frac{\mathcal{P}( m_{\mathrm{break}} | m_{\min }, m_{\mathrm{break}},-\alpha_1)}{\mathcal{P}\left(m_{\mathrm{break}} | m_{\mathrm{break}}, m_{\max },-\alpha_2\right))} \mathcal{P}(m_{1} | b, m_{\max },-\alpha_2)].
\end{split}
\end{equation}

\bibliography{ref}{}
\bibliographystyle{aasjournal}

\end{document}